\documentclass[11pt]{article}
\usepackage{moriond,epsfig}

\def\PRD{{\em Phys. Rev.} D}

\def\be{\begin{equation}}
\def\ee{\end{equation}}
\def\bea{\begin{eqnarray}}
\def\eea{\end{eqnarray}}

\def\W{W }
\def\Z{Z }
\def\t{t-quark }
\def\Wg{W$\gamma$ }
\def\Zg{Z$\gamma$ }
\def\WW{WW }
\def\WWZ{WWZ }
\def\WWg{WW$\gamma$ }

\def\D0{D0 }
\def\Etg#1{$E_{T}^{\gamma} > #1$\,GeV}
\def\Et#1{$E_{T} > #1$\,GeV}
\def\met#1{$\not{\! \! \! E_{T}} > #1$\,GeV}
\def\Met{$\not{\! \! \! E_{T}}$ }
\def\pt#1{$p_{T} > #1$\,GeV/$c$ }
\def\drlg#1{$\Delta R_{\rm \ell\gamma} > #1$ }
\def\pb{pb\,$^{-1}$ }

\begin{document}

%\vspace*{-0.5in}
%\begin{flushright}
%CDF/PUB/ELECTROWEAK/PUBLIC/6994 \\
%\today\\
%Version 1.1\\
%For submission to Proceedings of XXXIXth Rencontres de Moriond\\
%QCD and Hadronic Interactions, La Thuile, 28 March -- 4 April 2004.
%\end{flushright}
%\vspace*{-2cm}

\vspace*{4cm}
\title{DIBOSON PRODUCTION CROSS-SECTIONS AT $\sqrt{s}=1.96$\,TeV}
\author{ AIDAN ROBSON \\ on behalf of the CDF and D0 Collaborations }
\address{Department of Physics, Keble Road,\\
Oxford OX1 3RH, UK}

\maketitle\abstracts{
Recent results of \Wg, \Zg and \WW cross-section measurements
in the electron and muon channels are reported from p$\bar{\rm p}$
collisions at $\sqrt{s}=1.96$\,TeV recorded by the CDF and 
D0 collaborations.  Total cross-sections and kinematic distributions 
are found to be consistent with Standard Model expectations.}

\section{Introduction}

Diboson cross-section measurements provide tests of the
Standard Model and in particular a way of studying boson self-couplings.
The three groups of analysis presented here, \Wg, \Zg and WW, 
are complementary in the sense of being rather different from each other
 statistically, in their backgrounds, and in their leading systematic uncertainties.
From the point of view of studying boson self-couplings the analyses also
complement one another.  Whereas examining WW production 
probes the \WWZ and \WWg vertices, which are difficult to separate
experimentally, measuring \Wg production probes only the \WWg vertex.

In addition to the intrinsic interest of the measurements,
a solid understanding of diboson production is also important
for other physics: in t$\bar{\rm t}$ measurements if the Ws from the 
\t both decay leptonically the signature is very similar to a \WW event
as identified in these analyses;
 and a heavy Higgs would favour decay to two heavy bosons.

Inclusive \W and \Z cross-section measurements from CDF and D0
are reported elsewhere in these proceedings \cite{gavin}.
The recent preliminary \W and \Z cross-sections from CDF 
are precise to 2\% (neglecting the common luminosity error), 
which alongside their excellent agreement with NNLO predictions
provides the basis of our understanding for measuring the diboson
cross-sections.
The results presented here are very recent, and
in particular the \Wg results from D0 and the \WW results
from CDF are new for this conference.

\section{${\bf \rm p \bar{p} \rightarrow W \gamma \rightarrow \ell \nu \gamma}$ cross-section \label{wgsection}}
CDF and D0 report preliminary measurements of the \Wg production
cross-section using electron and muon decay channels.
In each case the initial selection of the W follows closely
that of the inclusive cross-section measurement.
CDF selects electron candidates from the central region of
the detector ($|\eta |<1.1$) that are isolated in the calorimeter
and have \Et{25}.  Requirements are made on the amount of
hadronic energy associated with the electromagnetic cluster,
on the electromagnetic shower shape, and on the quality and
matching of the associated track.
Muon candidates are selected in $|\eta |<1.1$ and are 
required to be isolated and to have small amounts of energy deposited
in the calorimeters.  The muon candidate tracks are required to 
have \pt{20}, to be well-matched to hits in the muon chambers,
and to be inconsistent with cosmic rays.
In addition to the lepton selection, candidate events
are required to have \met{20} in the muon channel, and \met{25}
in the electron channel.  A requirement is made on the transverse mass 
$m_{T({\rm \ell},\nu)}$, and to reject Zs a veto is made on 
second high-$p_T$ tracks in muon events.

The \Wg cross-section must be defined in a particular kinematic
region, and CDF chooses \Etg{7} and a separation between lepton
and photon \drlg{0.7}, where
$(\Delta R)^2 = (\Delta \eta)^2 + (\Delta \phi)^2$.
The photon selection is similar to the electron selection but
makes further use of the electromagnetic shower profile information,
and in addition to requiring calorimeter isolation a limit is made on
the number and energy of charged tracks pointing towards the cluster.
The fake photon background is large and understanding 
the probability for a jet to fake a photon is one 
of the principal challenges of this measurement.

In 202\pb CDF finds 259 events in the combined electron
and muon channels, compared with a Standard Model expectation
of $255.6\pm 2.1_{\rm stat}\pm 26.4_{\rm sys}$ events.
The preliminary measured cross-section is 
$\sigma\cdot Br({\rm W \gamma \rightarrow \ell \nu \gamma}) = (19.7 \pm 1.7_{\rm stat} \pm 2.0_{\rm sys} \pm 1.1_{\rm lum})$\,pb, to be compared with a
NLO prediction \cite{baur} of ($19.3\pm 1.4$)\,pb.

\begin{figure}
\begin{center}
%\vspace*{-5cm}
\epsfig{figure=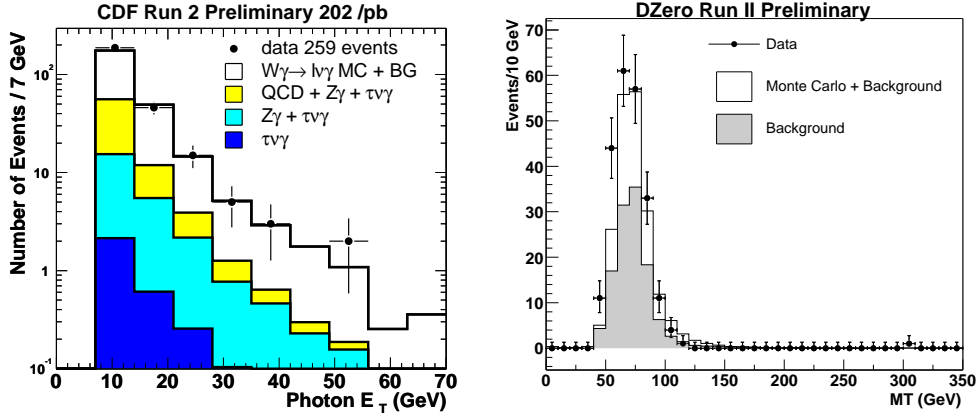,height=2.3in}
%\hspace*{-5cm}
\epsfig{figure=d0new_MT_rot.epsi,height=2.3in}
\caption{(left) $E_{T}^{\gamma}$ for CDF's \Wg candidate events; (right) 
$m_{T({\rm \ell},\nu)}$ for D0's \Wg candidate events.
\label{fig:cdfwg_phoet}}
\end{center}
\end{figure}

The $E_{T}^{\gamma}$ distribution is shown in Fig.~\ref{fig:cdfwg_phoet}.
A similar analysis, although with no $m_{T({\rm \ell},\nu)}$ requirement,
using Run\,I CDF data found an excess at high
$E_{T}^{\gamma}$ and it is interesting to see that no such excess is
observed in the current dataset.

The lepton selection used by D0 uses techniques similar to those
described above for the CDF analysis, although with an extended
coverage of $|\eta |<2.3$ for electrons and $|\eta |<2.0$ for muons.
Candidate electrons are required to have a cluster isolated in the 
calorimeter with high electromagnetic fraction and a matching track.
Muon candidates are required to be isolated in the calorimeter
and to have a central track matching a track in the muon spectrometer.
The photon is required to be central, $|\eta |<1.1$, and the total
$p_T$ of tracks pointing towards the photon candidate cluster is limited.
The measurement is made in the kinematic region \Etg{8} and \drlg{0.7}.
In total, 146 events are found in 162\pb of data in the 
electron channel compared with a Standard Model expectation of
$142\pm 17$ events, and 77 events are found in 82\pb of data in
the muon channel, compared with an expection of $67\pm 13$ events.
This leads to a preliminary cross-section measurement in the electron and muon channels of 
$\sigma\cdot Br({\rm W \gamma \rightarrow \ell \nu \gamma}) = (19.3 \pm 2.7_{\rm stat} \pm 6.1_{\rm sys} \pm 1.2_{\rm lum})$\,pb, to be compared with a
NLO prediction \cite{baur}  of ($16.4\pm 0.4$)\,pb.
The $m_{T({\rm \ell},\nu)}$ distribution of the candidate events
 is shown in Fig.~\ref{fig:cdfwg_phoet}.

\section{${\bf \rm p \bar{p} \rightarrow Z \gamma \rightarrow \ell \ell \gamma}$ cross-section}
CDF reports a preliminary measurement of the \Zg production 
cross-section using around 200\pb of data.
The first leg of \Z candidates is selected identically to the
lepton in the \Wg analysis described above.  Somewhat looser requirements
are made on the second leg, and the coverage of the analysis
is extended by allowing a second electron leg to be in the
forward regions of the detector, $1.2<|\eta|<2.8$, in which
case a matching silicon track is required.  An invariant mass
cut $m_{\ell \ell}>40$\,GeV/$c^2$ is applied, and in the case of
two central electrons or two muons, an opposite-charge requirement
is made on the tracks.
A photon is selected in the same way as for the \Wg analysis, and
as in the \Wg measurement the photon background has a large
uncertainty.  However \Z selection has much smaller background
than \W selection, and so the total background in the \Zg analysis
is small.  
In 202\pb of data, 69 \Zg candidate events are found compared with
a Standard Model expectation of $70.5\pm 4.0$ events.
The preliminary measured cross-section is 
$\sigma\cdot Br({\rm Z \gamma \rightarrow \ell \ell \gamma}) = (5.3 \pm 0.6_{\rm stat} \pm 0.3_{\rm sys} \pm 0.3_{\rm lum})$\,pb, to be compared with a
NLO prediction \cite{baur} of ($5.4\pm 0.3$)\,pb.  Kinematic distributions for the
\Zg candidates, for example the photon $E_T$ spectrum given in Fig.\,\ref{fig:cdfzgww},
show good agreement between data and Standard Model expectation.

\section{${\bf \rm p \bar{p} \rightarrow WW \rightarrow \ell \nu \ell \nu}$ cross-section}
CDF reports two measurements of the \WW cross-section: the first
making tight lepton cuts aiming for high purity, and the second
with a rather more open acceptance with the aim of increasing
statistics.

The `tight lepton' analysis selects two leptons according to 
criteria similar to those described in Section\,\ref{wgsection},
except with a universal $E_{T}$ cut of 20\,GeV.  Additional 
muon coverage is included by allowing second muon candidates that
do not point towards muon chambers and thus could not have a matching stub.
Leptons are required to have opposite charge, and candidate events
to have \met{25}.
To reduce fake \Met from mismeasured leptons, in events with lower
\Met, an angular separation is required between the \Met and the leptons. 
If same-flavour lepton pairs fall within an invariant mass window
corresponding to the \Z, an additional \Met significance cut is made.
Events having a jet with \Et{15} are rejected.
Although this increases the purity of the sample it does introduce
a systematic uncertainty in the acceptance as measured using simulation, 
as jet multiplicity is not particularly well-modeled in generators.
In 184\pb, 17 events are observed, with a signal to background
ratio of $\sim 2.3$, compared with a Standard Model
prediction of $16.1\pm 1.6$ events.  The preliminary resulting cross-section is
$\sigma\cdot Br({\rm p\bar{p} \rightarrow WW}) = (14.2 ^{+5.6}_{-4.9 {\rm stat}} \pm 1.6_{\rm sys} \pm 0.9_{\rm lum})$\,pb, to be compared with a
NLO prediction \cite{campbellandellis} of ($12.5\pm 0.8$)\,pb.

\begin{figure}
\begin{center}
\psfig{figure=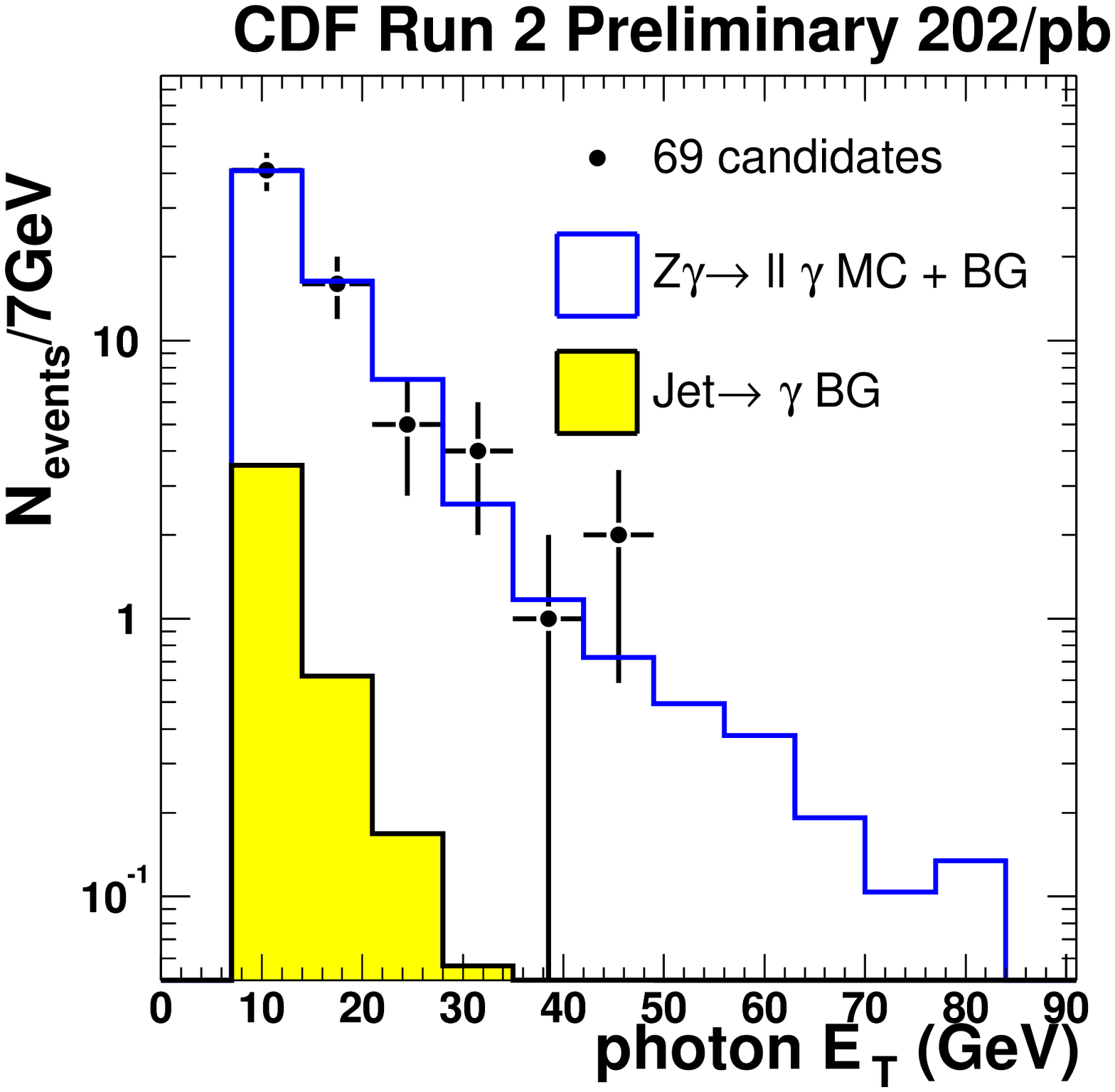,height=2.3in}
%\hspace*{1cm}
\psfig{figure=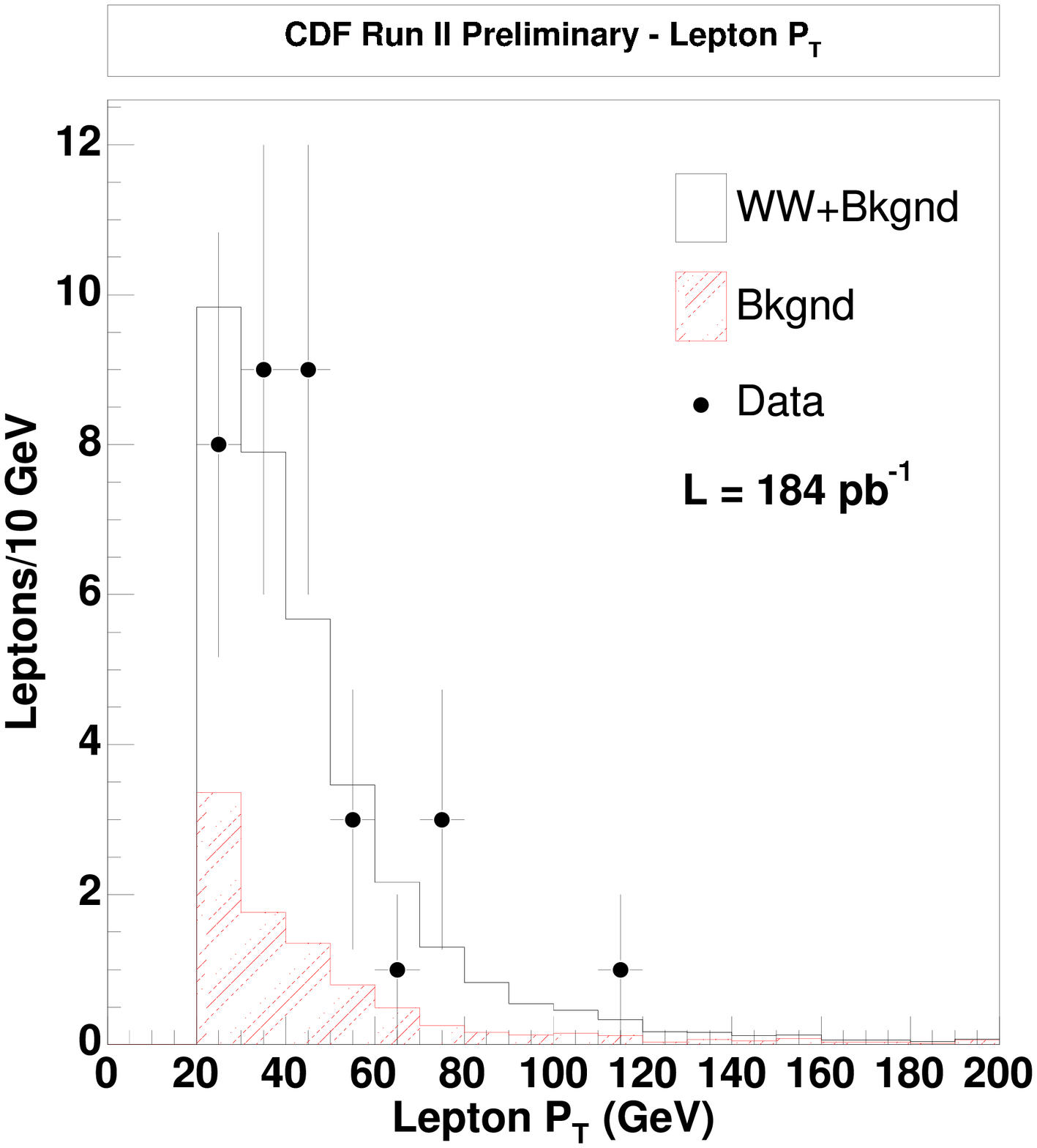,height=2.6in}
\caption{(left) $E_{T}^{\gamma}$ of CDF's \Zg candidate events; (right) 
Lepton $p_{T}$ spectra for CDF's \WW candidate events (tight lepton analysis).
\label{fig:cdfzgww}}
\end{center}
\end{figure}

The second analysis selects a tight electron or muon having \Et{20} 
as previously described, but makes the looser requirement of only an isolated
track for the second leg, increasing the acceptance compared with the
`tight lepton' analysis and giving some sensitivity to taus, but introducing
a large uncertainty associated with the fake-rate.
Events with more than one jet having \Et{20} are rejected, but single-jet
events are allowed.
Lepton candidates are required to have opposite charge, a requirement
is made on the angle between the \Met and the lepton candidates, and in the
\Z mass window an extra requirement is placed on the \Met significance.
39 events are selected in 200\pb with a signal to background
ration of $\sim 1.1$, compared with a Standard Model
prediction of $31.5\pm 1.0$ events.  The preliminary resulting cross-section is
$\sigma\cdot Br({\rm p\bar{p} \rightarrow WW}) = (19.4 \pm 5.1_{\rm stat} \pm 3.5_{\rm sys} \pm 1.2_{\rm lum})$\,pb, to be compared with a
NLO prediction of ($12.5\pm 0.8$)\,pb.
The lepton $p_{T}$ spectrum for the `tight lepton' analysis is shown in Fig.\,\ref{fig:cdfzgww}.

\section{Anomalous Coupling Limits.}
Work is in progress on the extraction of anomalous coupling
limits from the measured diboson cross-sections.  
Although LEP data were used to set good limits on anomalous couplings, 
the Tevatron can produce some events at much higher invariant mass
than at LEP and also can probe the \WWZ vertex alone, and thus
has rather different sensitivity.
\vspace*{0.35cm}
\begin{figure}[h]
\begin{minipage}[l]{0.4\textwidth}
\begin{center}
%\psfig{figure=xs_summary_win04.eps,height=2.7in}
	%\resizebox{0.4\textwidth}{!} { 
\includegraphics[height=3.in]{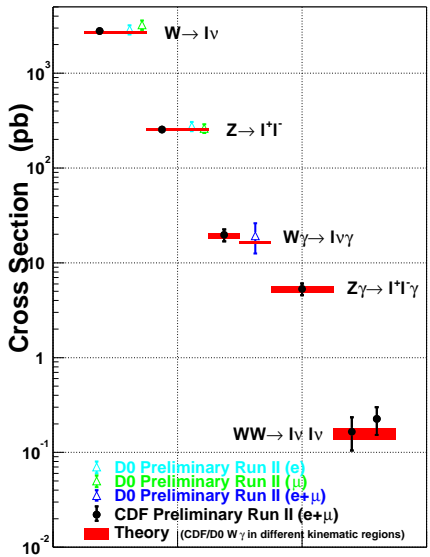}
	%}
\caption{Winter 2004 electroweak cross-section measurements from CDF and D0.
\label{fig:xssum}}
\end{center}
\end{minipage}
\hfill
\begin{minipage}[r]{0.57\textwidth}
\section{Conclusions}
The electroweak cross-section measurements from the Tevatron
for Winter 2004 are summarised in Fig.\,\ref{fig:xssum}.
Diboson cross-section measurements are getting
to the point where they are precision measurements;
measurements have been made of the \Wg, \Zg and \WW cross-sections
in the electron and muon channels, and all results are consistent
with Standard Model expectations.  Anomalous coupling limits will follow.

\end{minipage}
\end{figure}


\section*{References}
\begin{thebibliography}{99}

%\bibitem{an}ANOther{\it et al}, \Journal{\PRD}{99}{123}{2005}.
\bibitem{gavin}Gavin Hesketh, these proceedings.
\bibitem{baur} U.\,Baur {\em et al.}, \PRD\, {\bf 48} 5140 (1993).
\bibitem{campbellandellis}J.\,M.\,Campbell, R.\,K.\,Ellis, \PRD\, {\bf 60} 113006 (1999).

\end{thebibliography}
\end{document}